\documentclass[aps,pre,reprint,floatfix,superscriptaddress]{revtex4-1}
\pdfoutput=1
\usepackage{amsmath}
\usepackage{amssymb}
\usepackage{graphicx}
\usepackage[colorlinks=true,allcolors=blue]{hyperref}

\newcommand{\ipfnaffiliation}{Instituto de Plasmas e Fus\~{a}o Nuclear,
Instituto Superior T\'{e}cnico, Universidade de Lisboa, 1049-001 Lisboa,
Portugal.}
\newcommand{\polimiaffiliation}{Dipartimento di Energia, Politecnico di
Milano, Via Ponzio 34/3, 20133 Milan, Italy.}
\newcommand{\icreaaffiliation}{ICREA, Pg. Llu\'{i}s Companys 23, 08010
Barcelona, Spain.}
\newcommand{\ccfeaffiliation}{CCFE Culham Science Centre, Abingdon,
OX14~3DB, United Kingdom.}

\newcommand{\cS}{c_\text{S}}
\newcommand{\vA}{v_\text{A}}
\newcommand{\alfven}{Alfv\'{e}n}
\newcommand{\wA}{\omega_\text{A}}
\newcommand{\wGAM}{\omega_\text{GAM}}
\newcommand{\wTAE}{\omega_\text{TAE}}
\newcommand{\wBAAE}{\omega_\text{BAAE}}

\begin{document}
\title{High-order coupling of shear and sonic continua in JET plasmas}
\date{\today}

\author{Paulo~Rodrigues}
\affiliation{\ipfnaffiliation}
\author{Duarte~Borba}
\affiliation{\ipfnaffiliation}
\author{Francesca~Cella}
\affiliation{\ipfnaffiliation}
\affiliation{\polimiaffiliation}
\author{Rui~Coelho}
\affiliation{\ipfnaffiliation}
\author{Jorge~Ferreira}
\affiliation{\ipfnaffiliation}
\author{Ant\'{o}nio~Figueiredo}
\affiliation{\ipfnaffiliation}
\author{Mervi~Mantsinen}
\affiliation{\icreaaffiliation}
\author{Fernando~Nabais}
\affiliation{\ipfnaffiliation}
\author{Sergei~Sharapov}
\affiliation{\ccfeaffiliation}
\author{Paula~Sir\'{e}n}
\affiliation{\ccfeaffiliation}
\author{JET Contributors}
\affiliation{See author list of
\href{https://doi.org/10.1088/1741-4326/ac47b4}{J. Mailloux et al.,
Nucl. Fusion \textbf{62}, 042026 (2022)}.}

\begin{abstract}
A recent model coupling the shear-\alfven{} and acoustic continua, which
depends strongly on the equilibrium shaping and on elongation in
particular, is employed to explain the properties of \alfven{}ic
activity observed on JET plasmas below but close to the typical
frequency of toroidicity-induced \alfven{} eigenmodes (TAEs). The
frequency gaps predicted by the model result from high-order harmonics
of the geodesic field-line curvature caused by plasma shaping (as
opposed to lower-order toroidicity) and give rise to high-order geodesic
acoustic eigenmodes (HOGAEs), their frequency value being close to
one-half of the TAEs one. The theoretical predictions of HOGAE frequency
and radial location are found to be in fair agreement with measurements
in JET experiments, including magnetic, reflectometry and soft x-ray
data. The stability of the observed HOGAEs is evaluated with the linear
hybrid MHD/drift-kinetic code \texttt{CASTOR-K}, taking into account the
energetic-ion populations produced by the NBI and ICRH heating systems.
Wave-particle resonances, along with drive/damping mechanisms, are also
discussed in order to understand the conditions leading to HOGAEs
destabilization in JET plasmas.
\end{abstract}

\maketitle

\section{Introduction}

The stability of \alfven{} eigenmodes (AEs) is a key issue in
magnetically confined fusion, both for currently operating machines
(ASDEX-Upgrade, DIII-D, JET, etc) and for next-step devices such as
JT60-SA, being one of the major concerns regarding ITER
operation~\cite{fasoli.2007}. Once destabilized, AEs may degrade the
confinement of energetic ions produced by the heating systems (NBI,
ICRH) or by fusion reactions in burning plasmas and transport them
radially away from the core, compromising in this way the heating
process and eventually damaging the device walls or plasma facing
components~\cite{white.1995}.  Unstable AEs are known to arise in
frequency gaps of the MHD continuum that are produced by the coupling of
\alfven{} waves traveling along magnetic-field lines with a periodic
refractive index~\cite{zhang.2008}. The specific coupling mechanism
(equilibrium shaping, geodesic curvature, pressure) and the
dispersion-relation branches involved (shear-\alfven{} or slow acoustic,
etc) establish some of the AEs properties, like their frequency and
dominant polarisation, which are fundamental to understand their
interaction with the charged particles confined within the device and
the nature of the anomalous transport they may eventually induce.

Toroidicity-induced AEs (TAEs) with frequencies near the reference value
$\wTAE = \wA/(2q)$, where $\wA = \vA/R_0$ is the angular \alfven{}
frequency, $\vA$ is the \alfven{} speed, $R_0$ is the torus major
radius, and $q$ is the safety factor, are produced by the coupling of
two shear-\alfven{} (SA) waves and are one of the most extensively
studied \alfven{}ic instabilities in fusion devices due to their
potential to cause energetic-ion redistribution and
losses~\cite{fasoli.2007, gorelenkov.2014}. On the other hand, unstable
lower frequency AEs (i.e., with $\omega < \wTAE$) were first observed in
high-beta plasmas under intense NBI heating in
DIII-D~\cite{heidbrink.1993}, in association with high energetic-ion
loss levels much similar to those caused by TAEs~\cite{duong.1993}.
Since then, they have been documented in many other
devices~\cite{heidbrink.1999, heidbrink.2021}. Some properties of these
AEs have been explained by a MHD model coupling a single SA wave with
two acoustic waves via the lowest-order harmonic of the field-line
geodesic curvature due to toroidicity~\cite{holst.2000, gorelenkov.2007,
cheng.2019}, which results in two characteristic frequencies,
respectively
\begin{equation}
\begin{aligned}
  \wBAAE &= 2 \beta^\frac{1}{2} \, \wTAE
  \quad \text{and} \\
  \wGAM &= 2 \beta^\frac{1}{2} \sqrt{1 + 2 q^2} \, \wTAE.
\end{aligned}
\label{eq:wBAAE.and.wGAM}
\end{equation}
The first value corresponds to the top of the beta-induced acoustic AEs
(BAAEs) gap, with $\beta = \cS^2/\vA^2$ the plasma beta and $\cS$ the
sound speed, whilst the second corresponds to the bottom of the SA
continuum branch that is upshifted by finite pressure to a frequency
value that is typical of geodesic-acoustic modes (GAMs).  Alone or with
kinetic corrections, this model has been successfully employed to
explain experimental observations of AEs inside the frequency gaps near
$\wBAAE$~\cite{gorelenkov.2007b, gorelenkov.2009} and
$\wGAM$~\cite{heidbrink.2021}.
\begin{figure*}
  \begin{center}
  \includegraphics[width=510pt]{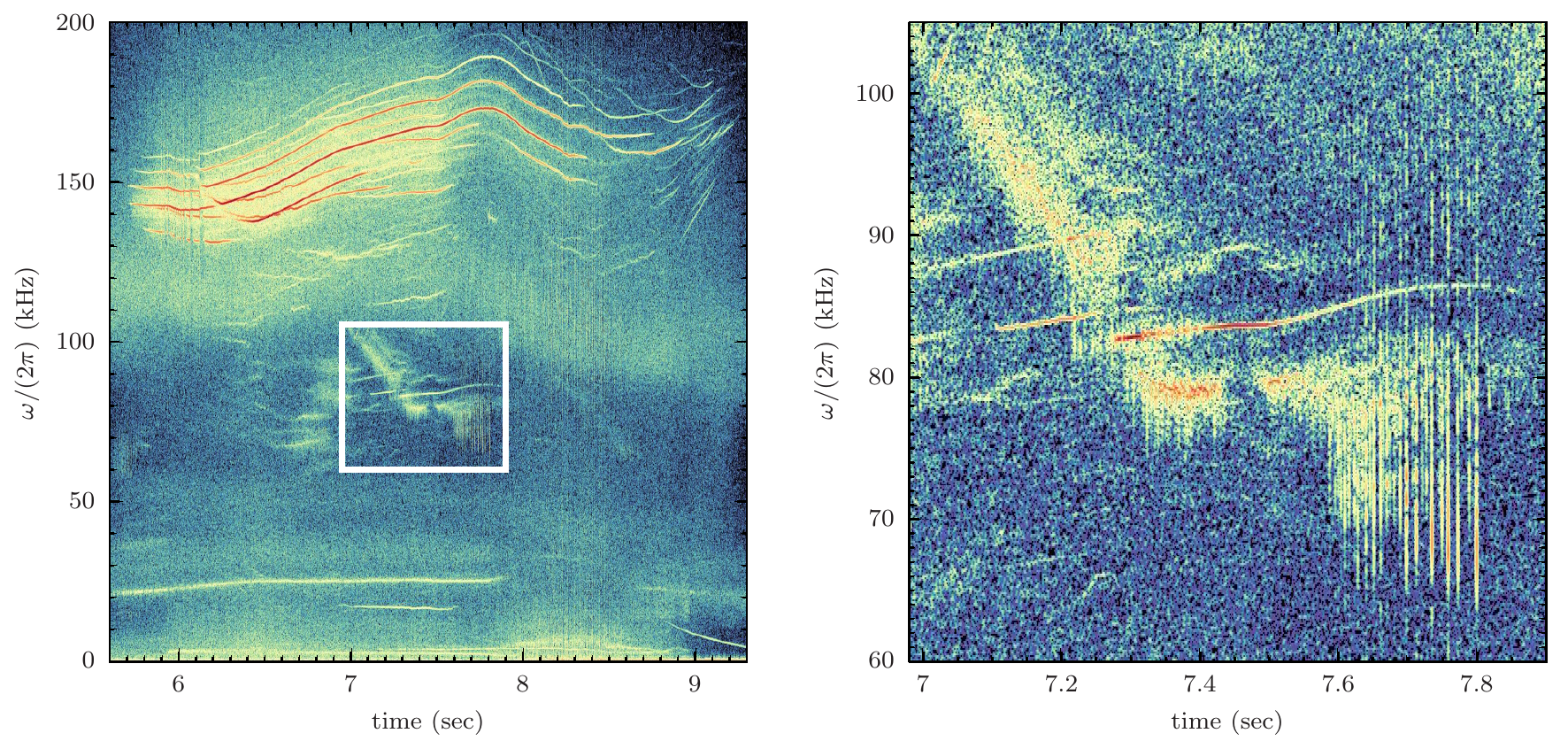}
  \end{center}
\caption{
\label{fig:spectrograms}
Magnetic-coil power spectra for JET pulse \#90199: ICRH-driven
\alfven{}ic activity, with TAEs around $150$~kHz and lower frequency AEs
in the range between $60$ to $100$~kHz (left); detail of the previous
spectrogram, evidencing the bursty frequency bands surrounding sharp
frequency lines (right).}
\end{figure*}

In this work, we report and discuss \alfven{}ic activity observed on JET
experiments, at about one-half of the frequency $\wTAE$, in plasmas
heated by NBI and ICRH. However, their properties (i.e., typical
frequency and radial location) are not easily described by the
aforementioned coupling model. In alternative, we show here that such
experimental measurements can be explained by a family of previously
unexplored continuum gaps that open when several acoustic waves couple
with a single SA wave via higher-order harmonics of the field-line
geodesic curvature (as opposed to lower-order toroidicity), which have
their origin on equilibrium shaping~\cite{rodrigues.2021}. In
particular, the finite elongation of JET plasmas is shown to play an
essential role in the argument. Although many families of gaps are
predicted by the analytic coupling model~\cite{rodrigues.2021}, some of
which have also been found by numerical approaches~\cite{huysmans.1995,
cheng.2019, kramer.2020}, here we focus on those gaps that open when
acoustic continuum branches are able to cross a SA branch whose bottom
has been lifted up to $\wGAM$. Therefore, the frequencies of interest
lie in the range $\wGAM \lesssim \omega \lesssim \wTAE$, that is,
significantly above the values predicted by previous lower-order
coupling models~\cite{holst.2000, gorelenkov.2007, cheng.2019} and
closer to the TAEs frequency. Although kinetic corrections are known to
change the \alfven{}-wave spectrum near the SA-continuum accumulation
point~\cite{zonca.1998} and produce kinetic eigenmodes with frequencies
above $\wGAM$~\cite{wang.2010, rizvi.2016}, this work is kept within the
framework of ideal MHD, with a perturbative drift-kinetic extension to
handle resonant wave-particle interactions.

The outline if this paper is as follows: In
section~\ref{sec:observations.and.models}, some of the lower-frequency
AEs most striking features are presented, along with the main
characteristics of the experimental scenarios where such observations
were made. The limitations of simpler low-order coupling models are
discussed, as well as the improved frequency predictions provided by the
more complete high-order model. In
section~\ref{sec:measurements.and.analysis}, a specific experimental
scenario is modelled numerically with an ideal-MHD code and the AEs
predicted frequency and radial location are compared with measurements.
These are found to agree with each other and with the analytical
estimates of the high-order coupling model.  In
section~\ref{sec:stability.and.resonances}, the resonant interaction
between low-frequency AEs and relevant plasma species (thermal ions and
energetic ions heated by NBI and ICRH) is addressed in its linear stage.
For the experimental scenario considered, the AEs stability is found to
be mainly established by the interplay between thermal-ion Landau
damping and the drive provided by the anisotropic population of
ICRH-accelerated energetic ions. The conclusions are summarised in
section~\ref{sec:discussion}.
\begin{figure}
  \begin{center}
  \includegraphics[width=246pt]{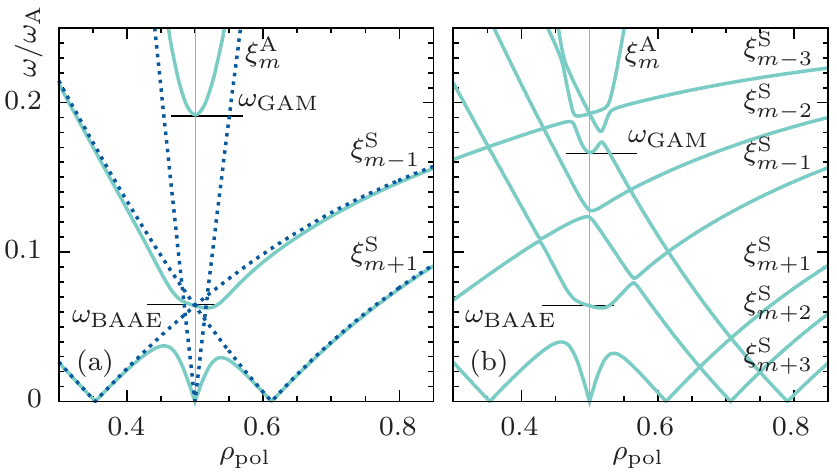}
  \end{center}
\caption{\label{fig:generic.continua}
Example continua of SA ($\xi^\text{A}_m$) and acoustic
($\xi^\text{S}_m$) harmonics of the plasma displacement
$\boldsymbol{\xi}$ for $q = 1 + 4 \rho_\text{pol}^2$, $\beta = 10^{-2}$,
$\varepsilon = 0.3$, poloidal mode number $m = 4$ and $n = 2$: (a)
uncoupled cylindrical-plasma limit (dots) and lowest-order coupling for
circular equilibria (solid lines); (b) high-order couplings enabled by
the elongation value $\kappa = 1.3$.}
\end{figure}

\section{
\label{sec:observations.and.models}
Experimental observations and continua-coupling models}

In order to optimise scenarios for the observation of TAEs driven
unstable by fusion $\alpha$-particles, experiments were conducted at JET
with deuterium plasmas that are characterised by low density, large core
temperature (usually associated with the presence of internal transport
barriers) and elevated safety factor~\cite{dumont.2018}. Injection of
ICRH power (hydrogen minority heating, concentrations in the range $2\%
\lesssim n_\text{H}/n_\text{e} \lesssim 7\%$, where $n_\text{e}$ is the
electron density) with the fundamental resonance located close to the
magnetic axis drives a large number of AEs unstable, as those seen in
figure~\ref{fig:spectrograms}.

Using reference values for these plasmas (on-axis field $B_0 \approx
3.4$~T, deuterium density $n_\text{D} = n_\text{e} \approx 4 \times
10^{19}$~m$^{-3}$, torus major radius $R_0 \approx 2.9$~m, and safety
factor $q \approx 1.5$, whence the values $\vA \approx 8.3 \times
10^6$~m/s and $\wA \approx 2.9 \times 10^6$~rad/s)~\cite{dumont.2018},
TAEs are expected to be found near $\wTAE/(2\pi) \approx 150$~kHz, this
frequency being shifted slightly upwards, according to their toroidal
mode number $n$, due to the Doppler shift caused by finite plasma
rotation. These estimates closely describe the TAEs seen in
figure~\ref{fig:spectrograms} around the predicted $150$~kHz value. In
the same spectrogram, additional activity is also seen at roughly one
half of the value corresponding to $\wTAE$, with well-defined frequency
lines surrounded by wide ($\sim 15$~kHz) frequency bands. As the
detailed plot in figure~\ref{fig:spectrograms} shows, these bands seem
to correspond to bursty events lasting a few milliseconds, in close
resemblance to those reported in recent DIII-D
observations~\cite{heidbrink.2021}. In contrast, however, the sharp
frequency lines between such bands is a distinctive feature of the AEs
being reported here. This feature seems to suggest that such AEs arise
inside well-defined frequency gaps and do not interact significantly
with the continuum nor are substantially affected by other sources of
strong damping.

The three-wave model (one SA, two acoustic)~\cite{holst.2000,
gorelenkov.2007, cheng.2019}, induced by finite toroidicity alone and
usually employed to describe continua coupling below $\wTAE$, is
insufficient to explain the AE frequencies in
figure~\ref{fig:spectrograms}. As illustrated by the frequency continua
plotted in figure~\ref{fig:generic.continua}(a) along the radial
variable $\rho_\text{pol} = (\Psi/\Psi_\text{b})^{1/2}$ (with $\Psi$ the
poloidal-field flux, $\Psi_\text{b}$ its value at the plasma boundary,
$\varepsilon = a/R_0$ the inverse aspect ratio, and $a$ the torus minor
radius) near a single resonant surface, the model predicts a single
frequency gap immediately below $\wBAAE$.  However, the latter's
relation with $\wTAE$ for a typical value $\beta \approx 10^{-2}$ sets
the upper limit for eventual BAAE frequencies to be around $30$~kHz, in
clear disagreement with the AEs being measured near $80$~kHz. On the
other hand, the same three-wave coupling model lifts the bottom of the
SA branch to $\wGAM \approx 70$~kHz. Besides the difference between
estimated and measured frequency values being still significant, it will
be shown in the next paragraph that equation~\eqref{eq:wBAAE.and.wGAM}
indeed overestimates the value $\wGAM$, which becomes lower if
plasma-shaping effects are taken into account.
\begin{figure}
  \begin{center}
  \includegraphics[width=246pt]{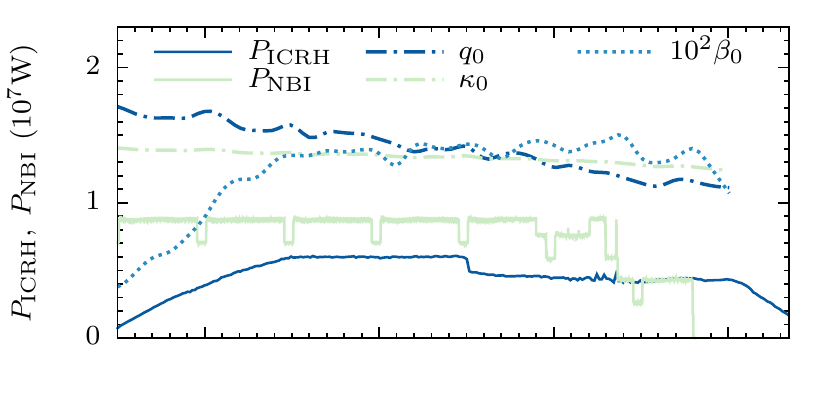}\\
  \vspace{-2em}
  \includegraphics[width=246pt]{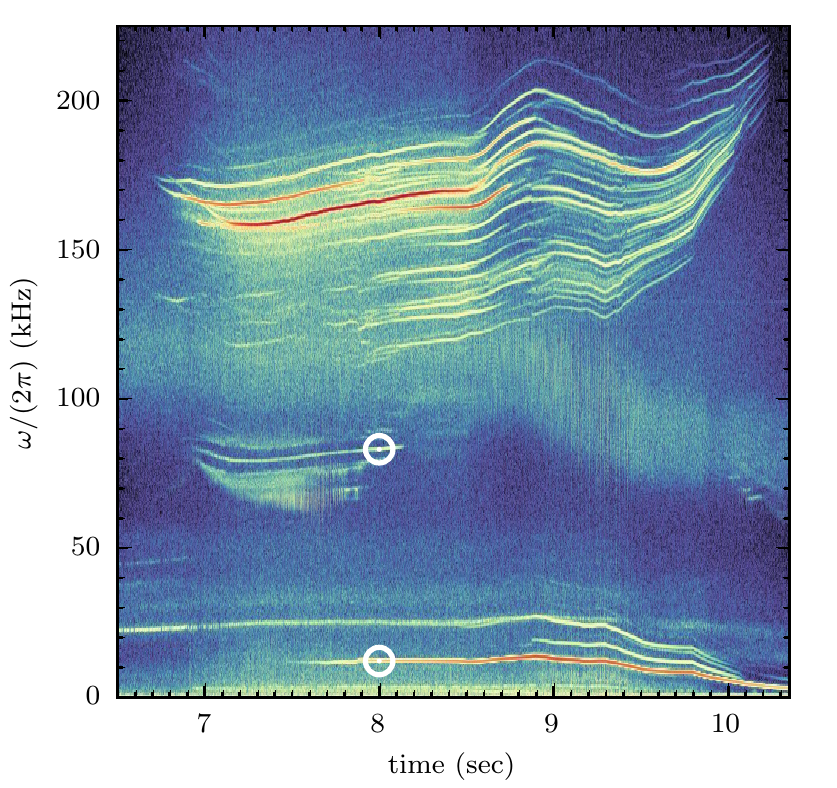}
  \end{center}
\caption{\label{fig:jet90198-data}
JET pulse \#90198: time evolution of injected power (ICRH and NBI) and
on-axis safety factor ($q_0$), elongation ($\kappa_0$), and plasma beta
($\beta_0$) values (top); Magnetic-coil power spectra showing a $n = 2$
HOGAE and a $3/2$ NTM (bottom). Open circles indicate the frequencies
measured at $8$ seconds.}
\end{figure}
\begin{figure*}
  \begin{center}
  \includegraphics[width=167pt]{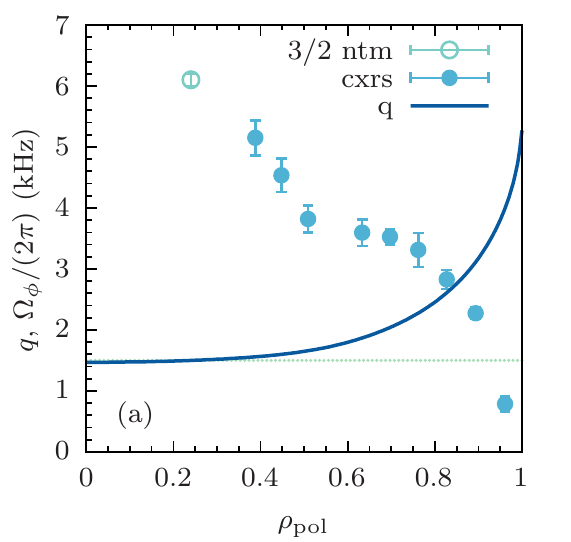}
  \includegraphics[width=167pt]{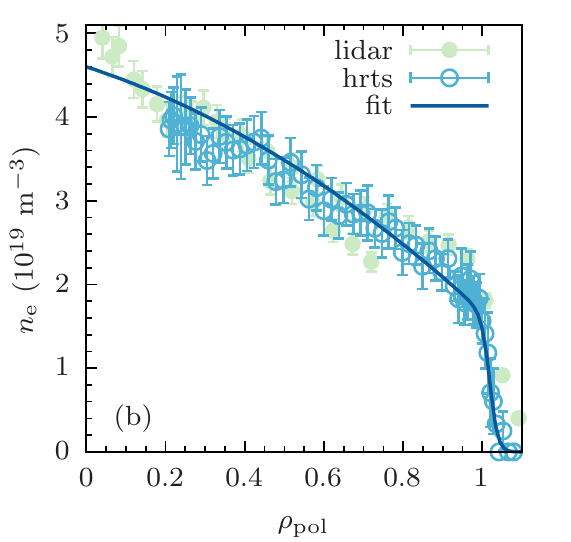}
  \includegraphics[width=167pt]{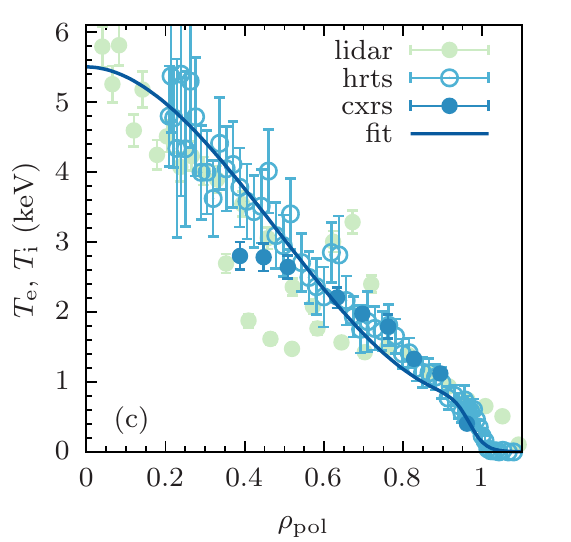}
  \end{center}
\caption{\label{fig:profiles}
Radial profiles for JET pulse \#90198 at 8 seconds: (a) toroidal
rotation frequency $\Omega_\phi$ data and safety factor (light
horizontal line at $q=3/2$); (b) fitted electron density $n_\text{e}$,
along with HRTS and LIDAR measurements; (c) fitted electron temperature
$T_\text{e}$ and available CXRS measurements of the ion temperature
$T_\text{i}$.}
\end{figure*}

An analytically tractable equilibrium model~\cite{rodrigues.2018} was
recently employed to show that plasma shaping, and in particular the
Shafranov shift and elongation, contribute with higher-order harmonics
to the geodesic curvature that couples SA and acoustic continua, in
addition to the lowest-order term arising from
toroidicity~\cite{rodrigues.2021}. As a consequence, the additional
couplings open frequency gaps where the acoustic branches are able to
cross the upshifted SA continuum, their frequencies being always in the
range $\wGAM \lesssim \omega \lesssim \wTAE$, as shown in
figure~\ref{fig:generic.continua}(b). Located near a rational surface $q
= m/n$ (rather than directly over it), with $m$ the poloidal mode
number, the largest frequency gap of these high-order family have its
width scaling as $\beta^{1/2}(\kappa^2 - 1)$, being thus closed for
circular plasmas where the elongation is $\kappa = 1$. Its frequency is
approximately given by~\cite{rodrigues.2021}
\begin{equation}
\omega = 2 \beta^\frac{1}{2} \wTAE \biggl[ 3 -
  \beta^\frac{1}{2} \sqrt{8 - 2 q^2} \Bigl(1 + \tfrac{3}{n q} \Bigr) +
    \cdots \biggr].
\label{eq:hogae.frequency}
\end{equation}
The toroidal mode number of the particular AE singled out in
figure~\ref{fig:spectrograms} is measured to be $n = 2$ by performing a
conventional spectral analysis over the data collected by a
magnetic-coil array distributed along the toroidal
direction~\cite{kim.1999}. Therefore,
equation~\eqref{eq:hogae.frequency} provides a frequency estimate of
$80$~kHz, which is in much better agreement with the measurements.

Other than opening additional frequency gaps, plasma shaping
(particularly elongation) also lowers the bottom of the SA continuum
from the value in equation~\eqref{eq:wBAAE.and.wGAM} down to the
corrected estimate~\cite{rodrigues.2021}
\begin{equation}
  \wGAM = \wBAAE \sqrt{1 + 2 q^2 \biggl(
    1 - \frac{3}{2} \frac{\kappa^2 - 1}{\kappa^2 + 1} + \cdots \biggr)},
\label{eq:wGAM.correction}
\end{equation}
as also depicted in figure~\ref{fig:generic.continua}. A reference
elongation value $\kappa \approx 1.5$ produces $\wGAM \approx 51$~kHz,
further distancing measurements and three-wave model predictions.
\begin{figure*}
  \begin{center}
  \includegraphics[width=510pt]{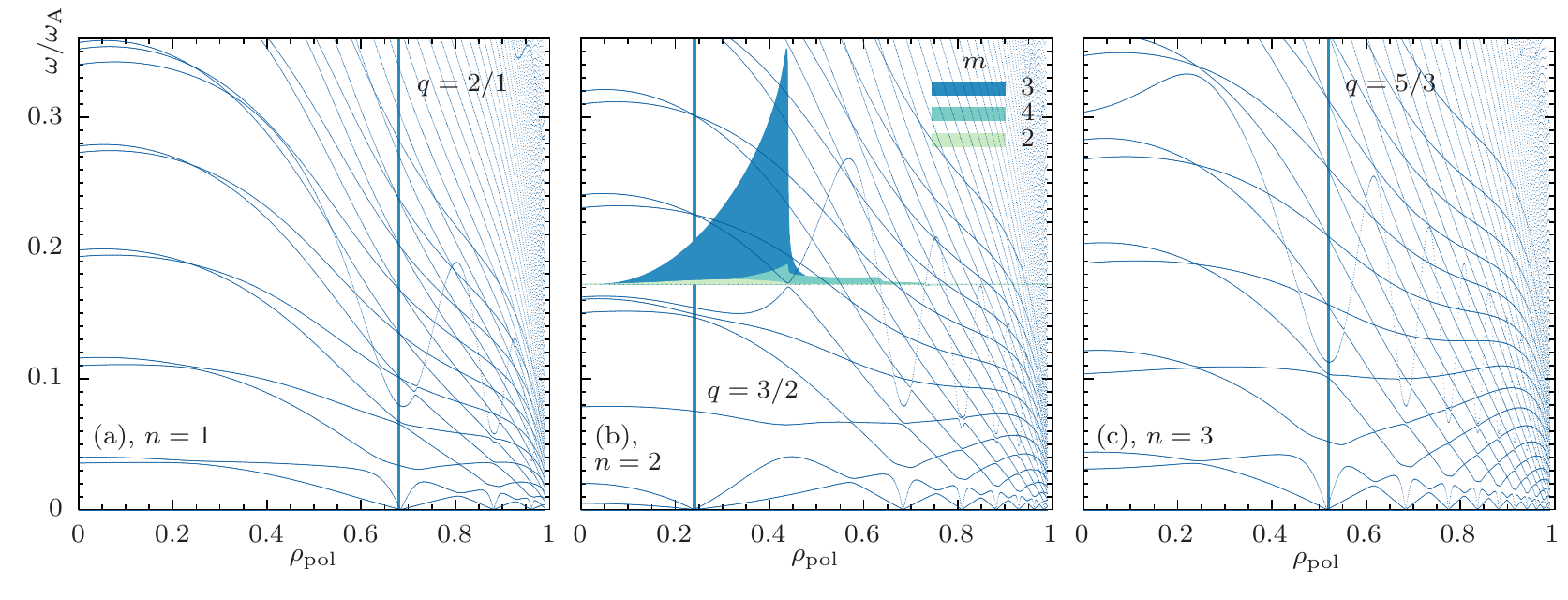}
  \end{center}
\caption{\label{fig:jet-continua}
Continuous spectra for JET pulse \#90198 at $8$~seconds and toroidal
mode numbers $n = 1$ (a), $2$ (b), and $3$ (c); vertical lines are
placed at the rational surfaces with lowest safety factor for each
toroidal mode number; radial structure of the $n = 2$ HOGAE (component
$q B_\phi^{-1} \boldsymbol{\xi} \cdot \nabla \rho_\text{pol}^2$,
dominant poloidal harmonics only, in a.u.), with the baseline matching
the value of its normalised frequency $\omega/\wA$.}
\end{figure*}

One important feature of the coupled SA-acoustic frequency gaps induced
by plasma shape is the existence condition~\cite{rodrigues.2021}
\begin{equation}
  q \leqslant
    \frac{2}{\sqrt{1 - \frac{3}{2} \frac{\kappa^2 - 1}{\kappa^2 + 1}}},
\label{eq:hogae.condition}
\end{equation}
involving the local safety factor and elongation. In short, a $q$ value
larger than $2$ prevents the gaps from opening, this limit being pushed
a bit upwards if $\kappa > 1$, which limits their existence to the inner
part of the plasma where the safety factor is usually lowest. In certain
circumstances, to be discussed later, high-order geodesic acoustic AEs
(HOGAEs) can be found within these gaps, which may account for the
properties of AEs below $\wTAE$ as those depicted in
figure~\ref{fig:spectrograms}.

\section{\label{sec:measurements.and.analysis}
Measurements and MHD analysis}

Unlike the spectrogram in figure~\ref{fig:spectrograms}, where multiple
frequency lines are seen within the range of interest (i.e., $\bigl[ 60;
100 \bigr]$~kHz), JET pulse \#90198 depicted in
figure~\ref{fig:jet90198-data} provides a simpler scenario for analysis:
it shows a clear $n = 2$ AE below the TAE group, with a single and very
sharp frequency line that is well separated from the two bursty bands
around it. At $8$ seconds, its frequency value is measured to be
$83.2$~kHz. At the same instant, a neoclassic tearing mode (NTM) with $m
= 3$ and $n = 2$ is visible at $12.2$~kHz, which can be used to estimate
the toroidal plasma-rotation frequency $\Omega_\phi = 6.1$~kHz at the
location of the $q = 3/2$ surface. The time trace in the top of
figure~\ref{fig:jet90198-data} shows the AE to arise with increasing
ICRH power and on-axis plasma beta, the latter remaining above the
reference value $10^{-2}$ while the eigenmode is visible. On-axis
elongation stays approximately constant and slightly below $1.4$,
whereas the safety factor drops steadily in time, being nonetheless
always below the critical value $2.8$ produced by
equation~\eqref{eq:hogae.condition}. From the same time trace, the AE
stabilisation at $8.1$~seconds cannot be related with decreasing ICRH
power (which provides the drive), plasma beta, or elongation (both
necessary to a non-vanishing gap width). Instead, it seems to be caused
by the safety-factor drop that moves the continuum outwards, as
explained later in this section

The magnetic equilibrium is reconstructed with
\texttt{EFIT}~\cite{lao.2005} and constrained by Faraday-rotation data,
yielding the safety factor profile depicted in
figure~\ref{fig:profiles}(a) that places the $q = 3/2$ surface at
$\rho_\text{pol} = 0.24$. This plasma rotation datum is also plotted in
figure~\ref{fig:profiles}(a), along with Charge eXchange Recombination
Spectroscopy (CXRS) measurements, thus providing a
$\Omega_\phi(\rho_\text{pol})$ profile that is employed to convert
between AE frequencies that are computed in the plasma frame and
measured Doppler-shifted frequencies. Other equilibrium parameters are
the magnetic axis location at $R_0 = 2.944$~m and the on-axis field $B_0
= 3.469$~T. Electron density $n_\text{e}(\rho_\text{pol})$ and
temperature $T_\text{e}(\rho_\text{pol})$ profiles are obtained by
fitting high-resolution Thompson scattering (HRTS) and LIDAR data, as
displayed in figures~\ref{fig:profiles}(b) and (c). Because no
thermal-ion density measurements are available, one assumes $n_\text{D}
= n_\text{e}$ and the plasma mass density being set by the Deuterium
ions alone.  Likewise, $T_\text{D} = T_\text{e}$ is also assumed,
following CXRS measurements for $\rho_\text{pol} \gtrsim 0.4$. Some
variations of these assumptions will be considered, later in
section~\ref{sec:stability.and.resonances}, in order to take into
account $Z_\text{eff}$ measurements and the uncertainties in the
$T_\text{D}$ profile.

\begin{figure}
  \begin{center}
  \includegraphics[width=246pt]{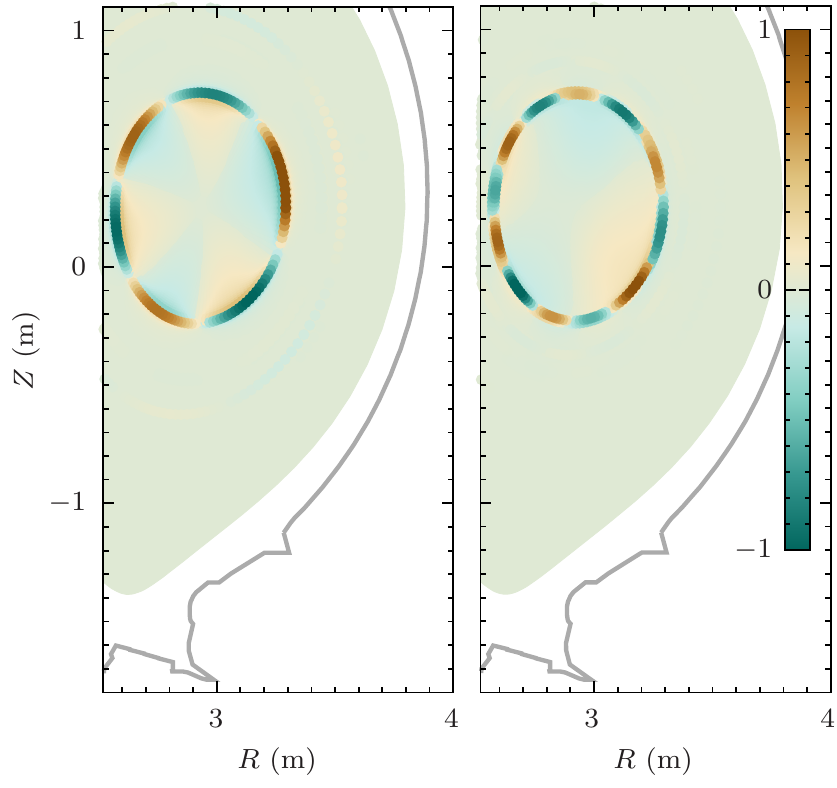}
  \end{center}
\caption{\label{fig:hogae-section}
Poloidal cross-section structure of the $n = 2$ HOGAE (JET pulse
\#90198, at $8$~seconds): normal electric-field perturbation
$\boldsymbol{E} \cdot \nabla \rho_\text{pol} / \bigl| \nabla
\rho_\text{pol} \bigr|$ (left) and parallel displacement
$\boldsymbol{\xi} \cdot \boldsymbol{B}^{(0)}/B^{(0)}$ (right).}
\end{figure}

The magnetic equilibrium, after refinement with the Grad-Shafranov
solver \texttt{HELENA}~\cite{huysmans.1991}, and the mass-density
profile are the inputs to the compressible ideal-MHD code
\texttt{CASTOR}~\cite{kerner.1998} and its continuous-spectrum extension
\texttt{CSCAS}~\cite{poedts.1993}. The latter is employed to produce the
three continua plotted in figure~\ref{fig:jet-continua} for the toroidal
mode numbers $n = 1, 2$, and $3$, while the former is used to compute
the frequency ($\omega/\wA = 0.173094$) and radial structure of an HOGAE
inside the high-order gap of the $n = 2$ continuum.  The coupled
SA-acoustic nature of the computed HOGAE is illustrated in
figure~\ref{fig:hogae-section}. The electric-field perturbation normal
to each flux surface is the plasma response to transverse (i.e.,
bi-normal) displacements and, therefore, shows a strong SA polarization
dominated by the same $m = 3$ harmonic $\xi^\text{A}_3$ seen in
figure~\ref{fig:jet-continua}.  Because this particular coupling is
induced by elongation, the dominant acoustic harmonic is
$\xi^\text{S}_{3 + 3}$~\cite{rodrigues.2021}, which is clearly visible
in the poloidal-section structure of the parallel displacement.
\begin{figure}
  \begin{center}
  \includegraphics[width=246pt]{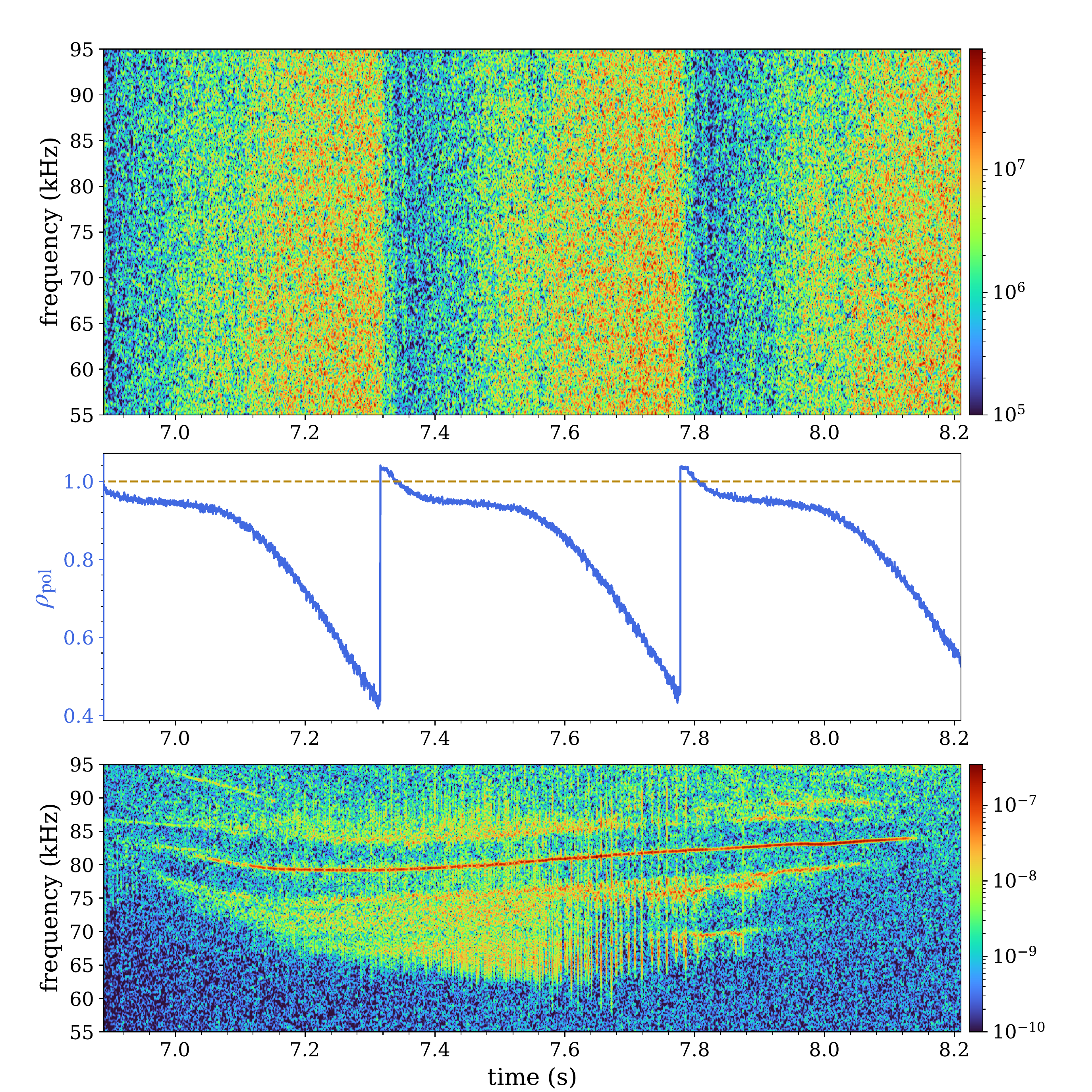}
  \end{center}
\caption{\label{fig:reflectometry}
Reflectometry data for JET pulse \#90198: reference spectrogram of the
magnetic-coil signal (bottom), radial location of the cut-off layer
(middle, blue), and reflectometer spectrogram (top), where no activity
is observed.}
\end{figure}

From the plots in figure~\ref{fig:jet-continua}, it is clear that only
for $n = 2$ can the HOGAE extend freely towards the magnetic axis
without crossing any continuum branch, therefore avoiding significant
continuum damping. This is so because HOGAE gaps are located at the
right-hand side of the upshifted SA continuum and a line of constant
frequency originating there will be blocked on its left-hand side if the
corresponding rational surface is too far away from the axis, as the $n
= 1, 3$ cases in figures~\ref{fig:jet-continua}(a,c) clearly illustrate.
The same blocking effect takes place if the safety factor decreases in
time, as seen in figure~\ref{fig:jet90198-data}, and rational surfaces
move outwards from the axis. Indeed, the HOGAE plotted in
figure~\ref{fig:jet-continua}(b) is just slightly above the continuum
near the axis, hinting that any further decrease of the safety factor
leads to its stabilisation by continuum damping.

Conversely, if the rational surface is sufficiently close to the axis,
as is the case for $n = 2$ in figure~\ref{fig:jet-continua}(b), the
continuum left-hand side stays beyond $\rho_\text{pol} = 0$ and an
eigenmode can be found through the gap. In view of this property,
experimentally observed HOGAEs with given toroidal and poloidal mode
numbers $n$ and $m$ can be regarded as proxies for the existence of a $q
= m/n$ rational surface close to the axis, which may be used to place
boundaries on the value $q(0)$.

The peculiar shape of the HOGAEs displacement harmonics, which tend to
concentrate towards the axis, makes them weakly sensitive to eventual
crossings with the continuum outwards from the gap radial location,
which for the $n = 2$ case in figure~\ref{fig:jet-continua}(b) is seen
to lie at $\rho_\text{pol} = 0.44$. Using the values for $R_0$ and $B_0$
returned by the equilibrium calculation and the on-axis Deuterium
density estimated from the fit to $n_\text{e}$ data in
figure~\ref{fig:profiles}(b), one gets the \alfven{} frequency on axis
$\wA/(2 \pi) = 426$~kHz. After correcting for the Doppler shift due to
$\Omega_\phi/(2 \pi) = 4.5$~kHz inferred at the gap radial location, the
HOGAE predicted frequency becomes thus $82.7$~kHz, which is fairly close
to the measured value that was reported at the beginning of this
section.

Many sources of experimental uncertainty may contribute to the
difference between predicted and measured frequency values, the most
important ones being concerned with the safety-factor and mass-density
profiles. While the former is closely related with the reliability of
equilibrium reconstruction performed by \texttt{EFIT}, the latter is
connected not only with eventual HRTS/LIDAR uncertainties but also with
the assessment of the several impurity fractions.

\section{\label{sec:constraints}
Eigenmode radial location and experimental constraints}

For the two JET pulses \#90198 and \#90199 considered in this work,
there are no reflectometry measurements available from the edge down to
the axis in order to unambiguously provide the AE location. As seen in
figure~\ref{fig:reflectometry} for the pulse \#90198 case, the
reflectometer cut-off layer is always restricted to the outer half of
the plasma (i.e., $0.5 < \rho_\text{pol} < 1$).  However, quite
significantly, no traces of the HOGAE can be found in the spectrogram of
the reflectometer signal (top) when it is compared with the one
corresponding to the magnetic coil (bottom). Therefore, the eigenmode
must be located in the inner half of the plasma ($\rho_\text{pol} <
0.5$), which agrees with the numerical solution computed by
\texttt{CASTOR} and plotted in figure~\ref{fig:jet-continua}(b).

\begin{figure*}
  \begin{center}
  \includegraphics[height=9.3cm]{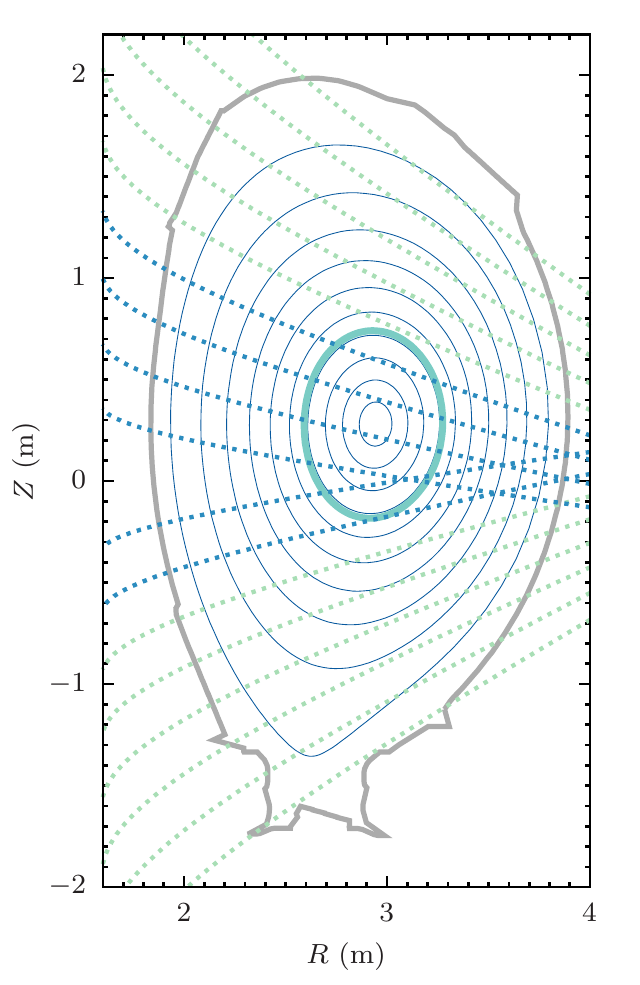}
  \hspace{1em}
  \includegraphics[height=9cm]{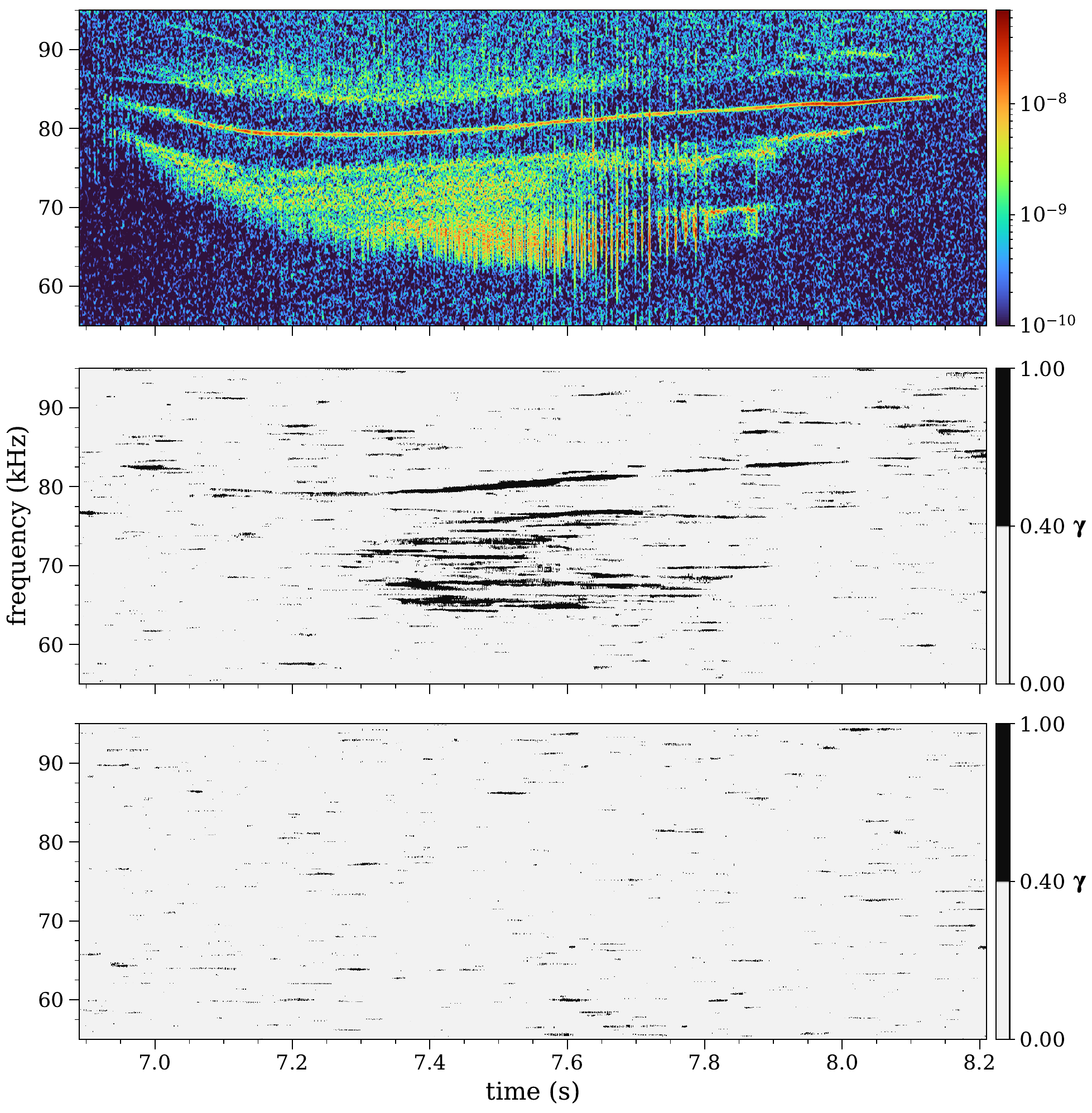}
  \end{center}
\caption{\label{fig:sxr}
SXR-coherence data for JET pulse \#90198: reference magnetic-signal
spectrogram (top right), phase-coherence between the magnetic signal and
a SXR channel crossing the AE (middle right) and for a channel not
crossing the AE (bottom right); SXR lines-of-sight of each channel
(left, dotted lines), flux surfaces at multiples of
$\Delta\rho_\text{pol} = 0.1$ (left, thin solid lines), and at the HOGAE
gap location (left, thick solid line); channels plotted with dark dotted
lines show statistically relevant coherence, those plotted with light
dotted lines do not.}
\end{figure*}

Although the HOGAE perturbation is too weak to show up clearly in soft
x-rays (SXR) signals, it is possible to find statistically significant
phase coherence between the latter and magnetic-coil signals for a
certain number of SXR channels as depicted in figure~\ref{fig:sxr}.
Indeed, all lines-of-sight that cross magnetic surfaces with
$\rho_\text{pol} < 0.4$ show some trace of the perturbation, similar to
the example provided (right-hand side, middle panel).  On the contrary,
all lines-of-sight lying entirely at $\rho_\text{pol} > 0.45$ do not
show any traces (right-hand side, bottom panel). Again, these
experimental constraints support the HOGAE radial structure depicted in
figure~\ref{fig:jet-continua}(b).

\section{\label{sec:stability.and.resonances}
Eigenmode stability and resonant interactions}

The resonant energy exchange between HOGAEs and an ion species $s$ is
evaluated by the hybrid MHD/drift-kinetic code
\texttt{CASTOR-K}~\cite{borba.1999, nabais.2015} following a
perturbative approach, which returns the linear growth rate as
\begin{equation}
\frac{\gamma_s}{\omega} =
  - \frac{1}{2 \omega^2} \int
    \text{Im} \bigl(L_{(1)}^\ast f_s^{(1)} \bigr) d^3x d^3v
  \bigg/
    \int \rho \boldsymbol{\xi} \cdot \boldsymbol{\xi}^\ast d^3x
\label{eq:growth.rate}.
\end{equation}
Here, $L_{(1)}$ and $f_s^{(1)}$ are the linear response of the
guiding-center Lagrangian and equilibrium distribution function
$f_s^{(0)}$ to the MHD displacement eigenfunction $\boldsymbol{\xi}$
computed by \texttt{CASTOR}. The integrals in
equation~\eqref{eq:growth.rate} are evaluated in the space of the
guiding-center constants of motion: energy $E$, toroidal momentum
$P_\phi$, and $\Lambda = \mu B_0/E$, with $\mu$ the magnetic
moment~\cite{porcelli.1994}.

\begin{figure}
  \begin{center}
  \includegraphics[width=246pt]{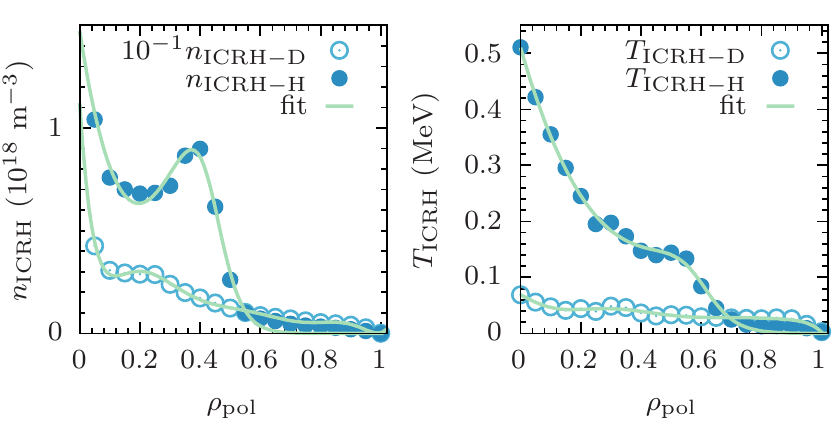}
  \end{center}
\caption{\label{fig:pion}
Density and temperature of ICRH-heated H and D ions computed by
\texttt{PION} for JET pulse \#90198 at $8$~seconds.}
\end{figure}

For such purpose, thermal-bulk Deuterium ions are described by a local
Maxwellian, their density and temperature being, for the moment, those
in figure~\ref{fig:profiles} (i.e., $T_\text{th-D} = T_\text{i} =
T_\text{e}$ and $n_\text{th-D} = n_\text{i} = n_\text{e}$, which
integrates radially to yield $N_\text{e} = 1.97 \times 10^{21}$
electrons within the plasma volume). The distribution of NBI-heated
Deuterons is computed by the heating code
\texttt{ASCOT}~\cite{hirvijoki.2014}.  In turn, the code
\texttt{PION}~\cite{eriksson.1993, mantsinen.1999} provides the
distributions of ICRH-heated H and D ions, their density and temperature
being plotted in figure~\ref{fig:pion}. The strong anisotropy of the hot
H ion population is modelled by a separable distribution
\begin{multline}
f_\text{ICRH-H}^{(0)} \propto n_\text{ICRH-H}(\rho_\text{pol}) \\
  \times \exp \biggl[ - \frac{E}{T_\text{ICRH-H}} \biggr]
    \exp \biggl[ - \frac{(\Lambda - \Lambda_\text{ICRH})^2}{
      2 \delta_\text{ICRH}^2} \biggr],
\label{eq:distro-icrh}
\end{multline}
where $2 R_0 \delta_\text{ICRH}$ is the Doppler broadening of the ICRH
resonant layer centred around the vertical line $R_\text{ICRH}/R_0 =
\Lambda_\text{ICRH}$. In the following, the fundamental ICRH resonance
of H ions is assumed to be located on axis, whence $\Lambda_\text{ICRH}
= 1$. On the other hand, ICRH-heated Deuterons are assumed to be already
thermalised, being thus also described by a local Maxwellian.

\begin{figure}
  \begin{center}
  \includegraphics[width=246pt]{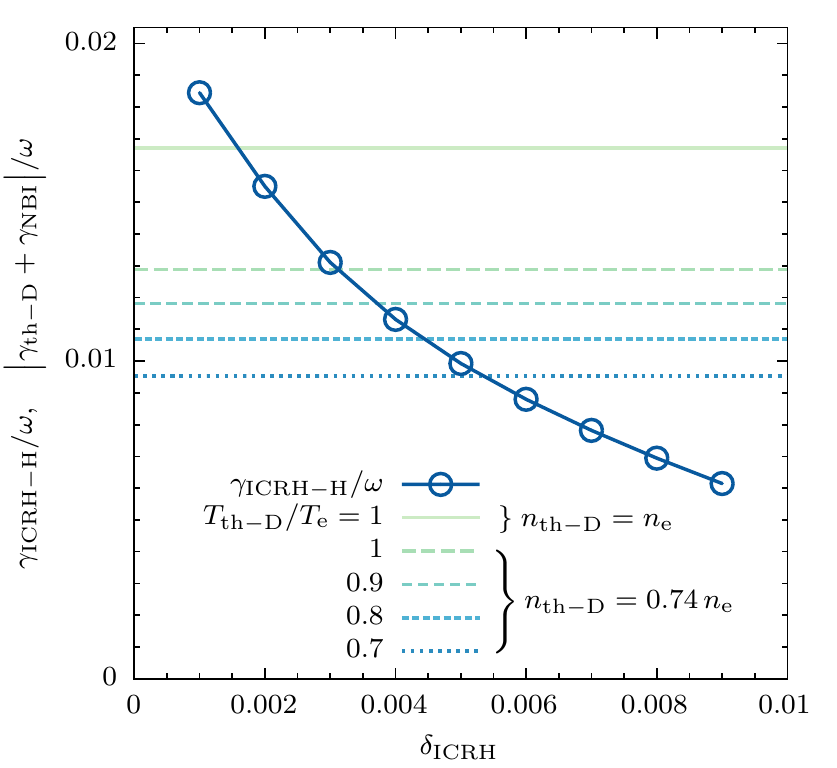}
  \end{center}
\caption{\label{fig:growth-rate}
Scan of ICRH-heated H ions driving rate over the normalised width
$\delta_\text{ICRH}$ of the resonance layer (open circles) and the
combined damping rate ($\bigl| \gamma_\text{th-D} + \gamma_\text{NBI}
\bigr|/\omega$, horizontal lines) under different assumptions regarding
the ratios $n_\text{th-D}/n_\text{e}$ and $T_\text{th-D}/T_\text{e}$.}
\end{figure}

The damping rates computed by \texttt{CASTOR-K} for the thermal and NBI
Deuterium ions are $\gamma_\text{th-D}/\omega = -0.015$ and
$\gamma_\text{NBI}/\omega = -0.002$, respectively. ICRH Deuterons have a
negligible damping contribution (of order $10^{-4}$) due to their modest
number ($N_\text{ICRH-D}/N_\text{e} \sim 10^{-2}$) and temperature
($T_\text{ICRH-D} \sim 50$~keV). In turn, the drive from ICRH-heated H
ions depends strongly on the gradient $\partial_\Lambda
f_\text{ICRH-H}^{(0)} \sim f_\text{ICRH-H}^{(0)}/\delta_\text{ICRH}$,
which is related with the distribution-function anisotropy, and thus
with the normalised Doppler broadening $\delta_\text{ICRH}$ of the
resonant layer. A scan of the driving rate $\gamma_\text{ICRH-H}/\omega$
is plotted in figure~\ref{fig:growth-rate} for a range of
$\delta_\text{ICRH}$ values, along with the combined damping of thermal
and NBI-heated Deuterons obtained under different assumptions in order
to understand their effect on the overall stability of the eigenmode.
Indeed, if $n_\text{th-D} = n_\text{e}$ and $T_\text{th-D} =
T_\text{e}$, the drive due to a $0.5\%$ fraction of ICRH-heated H ions
(as computed by \texttt{PION} with respect to the electron number) is
not sufficient to surpass the combined damping and drive the HOGAE
unstable, except for the lowest value of $\delta_\text{ICRH}$.  Besides
uncertainties in the fraction of hot H ions (on which the driving rate
depends linearly), other effects may contribute to a different balance
of the energy exchange.

One such effect is the thermal-ion dilution due to the many impurities
(in unknown concentrations) that are responsible to raise the effective
charge number to the measured value $Z_\text{eff} \sim 2.4$. To keep the
analysis simple, an impurity mix of $3\%$ Beryllium and $0.1\%$ of Neon
and Nickel is considered, along with $4\%$ of minority H, which keeps
the quasi-neutrality and $Z_\text{eff}$ consistent with measurements,
while diluting the thermal Deuterium down to $80\%$. The latter must be
further reduced by $6\%$, in order to account for the $3\%$ of Deuterons
heated by ICRH (as computed by \texttt{PION}) and $3\%$ by NBI (as
reported by \texttt{ASCOT}). Overall, one finds $n_\text{th-D} = 0.74
n_\text{e}$ and the thermal ion damping reduces accordingly, raising the
critical value of the resonance-layer width below which the HOGAE
becomes unstable. Another effect is caused by the uncertainty of the
profile $T_\text{th-D}(\rho_\text{pol})$, which is constrained by CXRS
data only in the outer half of the plasma (recall
figure~\ref{fig:profiles}). The impact of reducing the ratio
$T_\text{th-D}/T_\text{e}$ on the combined damping rate is depicted in
figure~\ref{fig:growth-rate}, where the critical resonance-layer width
is seen to increase further. Additional sources of damping, like
continuum or radiative damping, as well as the eventual impact of other
non-ideal-MHD effects, were not taken into account.

\begin{figure}
  \begin{center}
  \includegraphics[width=246pt]{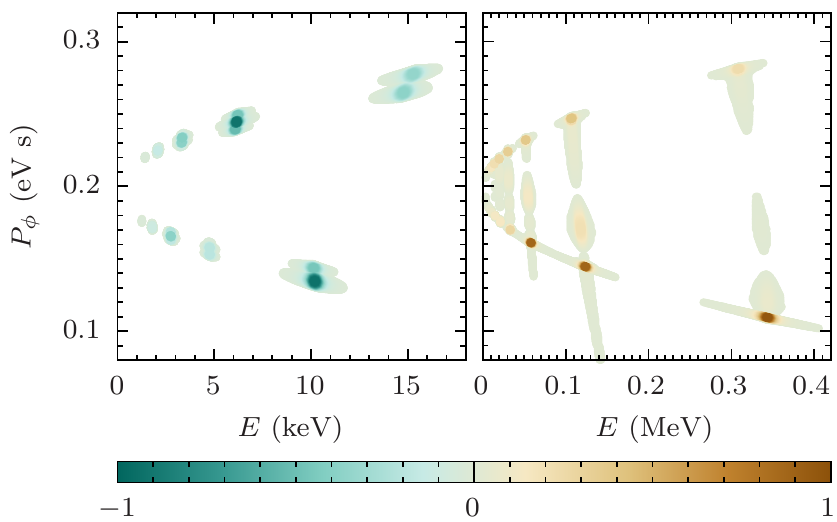}
  \end{center}
\caption{\label{fig:resonances}
Resonant energy exchange (shading code, a.u.): damping due to strongly
passing thermal D ions ($\Lambda \sim 0$, left) and drive due to trapped
hot H ions at the destabilising side of the anisotropic distribution
($\Lambda = \Lambda_\text{ICRH} + \delta_\text{ICRH}$, right).}
\end{figure}

The major wave-particle resonances are plotted in
figure~\ref{fig:resonances} for thermal Deuterons and ICRH-heated H
ions. The interactions of the former are dominated by strongly passing
particles ($\Lambda \sim 0$), whose stronger resonances are located at
$5$~keV and $10$~keV. These values are close to the plasma temperature
at the AE location but significantly below the fundamental resonance at
$v_\parallel \sim 3 \cS$~\cite{rodrigues.2021}, which corresponds
approximately to $50$~keV in this case. In turn, resonances with ICRH
hot H ions are stronger at $340$~keV, with two slightly less intense
peaks near $120$~keV and $60$~keV, all close to the H temperature
provided by \texttt{PION} (i.e., about $150$~keV).

\section{\label{sec:discussion} Discussion}

In summary, \alfven{}ic activity observed on JET experiments below but
close to the typical TAE frequency was reported. Its properties were
explained using a model coupling the SA and acoustic continua that
depends strongly on equilibrium shaping and, in particular, on the
elongation of JET plasmas. Predicted frequencies and radial locations of
the HOGAEs arising from such coupling in a specific JET discharge were
found to be in fair agreement with experimental measurements, which
include magnetic, reflectometry, and SXR data. The stability assessment
confirms HOGAEs instability in the presence of ICRH-heated ions with
energies around 340~keV, if the resonance layer is sufficiently narrow
or, correspondingly, if their distribution-function anisotropy is
sufficiently high.  The effects of thermal-ion dilution by impurities
and different $T_\text{i}/T_\text{e}$ ratios on the stability threshold
were also addressed.

Numerical results indicate that anisotropy is a good candidate to drive
HOGAEs unstable in the JET scenarios under analysis. Indeed,
figure~\ref{fig:growth-rate} hints that isotropic ICRH distributions
(i.e., in the limit $\delta_\text{ICRH} \rightarrow \infty$) are unable
to surpass thermal-ion Landau damping unless $T_\text{i}/T_\text{e}$ is
unrealistically low. However, this conclusion applies to the specific
ICRH population considered in this work, which is characterised by
temperatures below $200$~keV at the eigenmode location. The effect of
isotropic populations with much higher energy (e.g., fusion alphas) on
HOGAEs stability is still an open question.

\begin{acknowledgments}
This work has been carried out within the framework of the EUROfusion
Consortium, funded by the European Union via the Euratom Research and
Training Programme (Grant Agreement No.~101052200 - EUROfusion). Views
and opinions expressed are, however, those of the author(s) only and do
not necessarily reflect those of the European Union or the European
Commission. Neither the European Union nor the European Commission can
be held responsible for them. One of the authors (FC) was supported by
FuseNet from the Euratom research and training programme under Grant
Agreement No.~633053. IPFN activities were also supported by
“Funda\c{c}\~{a}o para a Ci\^{e}ncia e Tecnologia” (FCT) via projects
UIDB/50010/2020 and UIDP/50010/2020.
\end{acknowledgments}

\end{document}